\documentclass[11pt]{article}
\usepackage{epsfig}
\usepackage{graphicx}
\usepackage{amssymb}

\begin{document}

\title{A Michelson interferometer in the field of a plane gravitational wave}
\author{Nikodem J. Pop\l awski}
\date{}

\maketitle
\begin{center}
{\it Department of Physics, Indiana University, 727 East Third Street, Bloomington, Indiana 47405}\\
Email: nipoplaw@indiana.edu
\end{center}

\begin{abstract}

We treat the problem of a Michelson interferometer in the field of a plane gravitational wave in the framework of general relativity. The arms of the interferometer are regarded as the world lines of the light beams, whose motion is determined by the Hamilton-Jacobi equation for a massless particle. In the case of a weak monochromatic wave we find that the formula for the delay of a light beam agrees with the result obtained by solving the linearized coupled Einstein-Maxwell equations. We also calculate this delay in the next (quadratic) approximation.

\end{abstract}


\section{Introduction}

The subject of this work is to investigate the behavior of a Michelson interferometer in the presence of a plane gravitational wave. Sources of gravitational waves include the collapse of massive stars in supernova explosions, and the coalescence of stars in binary systems containing a neutron star and/or a black hole. The large distances from these objects allow us to assume that the gravitational waves detected on the Earth may be regarded as plane waves.

Gravitational waves were predicted by Einstein in his general theory of relativity~\cite{e1,e2} as perturbations of the spacetime metric that propagate at the speed of light. Plane gravitational waves were considered first by Rosen~\cite{1}, Taub~\cite{2}, and McVittie~\cite{3}, and their work seemed to indicate that such waves may not physically exist. However, Bondi, Pirani, and Robinson \cite{4,5,6} proved that this conclusion was wrong, and that these waves not only exist but also carry energy. Exact solutions of the Einstein equations, which describe plane gravitational waves, have also been developed in~\cite{ex1,ex2}. In 1975, Hulse and Taylor found indirect proof for the existence of gravitational waves by observing the pulsar PSR 1913+16~\cite{8}. They noticed that the orbital parameters of this pulsar were changing according to a model assuming that such a system radiates gravitational waves~\cite{8b,9}. This discovery was awarded with the Nobel Prize in physics in 1993.

Research on the possibility of detecting gravitational waves was started in the 1960s by Weber~\cite{7}, who constructed the first resonant-mass antenna. His results claiming the discovery of gravitational waves~\cite{7b} could not be confirmed by other groups who built similar detectors. In the 1970s, another method of detection was suggested using laser interferometry~\cite{For,RS,Est,ST}. Current and planned ground-based laser interferometric detectors include AIGO (Australia)~\cite{12}, GEO 600 (Germany/UK)~\cite{13}, LIGO (USA)~\cite{14}, TAMA 300 (Japan)~\cite{15}, and VIRGO (France/Italy)~\cite{16}. The first search for signals from neutron stars in LIGO and GEO 600 data did not find direct evidence for gravitational waves~\cite{res2}, but gave upper limits on the strength of periodic gravitational radiation~\cite{res1}. In 2011, ESA and NASA plan to launch LISA~\cite{17}, a space project that will be able to make observations in a low-frequency band that is not achievable by ground-based observatories. A bibliography on gravitational wave theory and experiment through 1999 is listed in~\cite{bibl}.

Since the search for gravitational waves is mainly based on interferometry, we consider a Michelson interferometer in the presence of a plane gravitational wave. The effect of such a wave on a light beam, using the linearized coupled Einstein-Maxwell equations, was studied in~\cite{lo,fr}. However, these equations in the quadratic approximation may lead to quite lengthy expressions. Instead, we can treat the light beam as photons whose motion is determined by the geodesic equations for a massless particle. 

The equations of motion for a particle in the metric of a strong gravitational wave~\cite{9} were derived by Ebner~\cite{eb}, and all seven integrals of motion can be found in the literature, e.g.~\cite{ab}. A simple model of a Michelson interferometer in the field of such a wave, based on the above equations of motion, was proposed by Ba\.{z}a\'{n}ski~\cite{11}. Here, we generalize this model to the second (quadratic) approximation. For simplicity, we consider a monochromatic plane gravitational wave whose direction is parallel to one arm of the interferometer and perpendicular to the other. Since the energy of gravitational waves contains the factor $\frac{G}{c^{5}}$, the amplitude for such waves is very small. Therefore, second-order effects in gravitational-wave detection are irrelevant experimentally, although it might be interesting to explore them.

This paper is organized as follows. In section~2 we introduce a plane gravitational wave and review the equations of motion for a test particle. In section~3 we study the behavior of a Michelson interferometer under the action of this wave in the first (linear) approximation. In section~4 we proceed to the second approximation. The results are summarized in section~5.
For the metric tensor and the curvature tensor we use the conventions of~\cite{9}.

\section{Motion of a particle in the field of a plane gravitational wave}

In this section we derive the motion of a test particle in the presence of a plane gravitational wave~\cite{eb,11} using the relativistic Hamilton-Jacobi equation. Let us consider the metric tensor whose components are functions of only one variable $u$, $g_{ab}=g_{ab}(u)$.
In this case, it can be shown~\cite{9} that the spacetime interval in vacuum is of the form
\begin{equation}
ds^{2}=du\,dv+g_{ab}(u)dx^{a}dx^{b},
\label{eq:int}
\end{equation}
where the letters $a,b$ refer to the coordinates $2,3$.
For a plane wave propagating in the direction of the $x$ axis we have $u=ct-x$ and $v=ct+x$. If the interval is given by equation~(\ref{eq:int}) then all components of the Ricci tensor identically vanish except
$R_{uu}=-\frac{1}{2}{\dot{\kappa}}^{a}_{a}-\frac{1}{4}\kappa^{a}_{b}\kappa^{b}_{a}$,
where $\kappa_{ab}={\dot{g}}_{ab}$,
and the dot denotes differentiation with respect to $u$.
Thus, the Einstein field equations in the case of a plane gravitational wave reduce to one equation, $R_{uu}=0$.

The Hamilton-Jacobi equation for a particle with mass $m$ moving in the gravitational field described by the metric in (\ref{eq:int}) is given by
\begin{equation}
4\frac{\partial S}{\partial u}\frac{\partial S}{\partial v}+g^{ab}\frac{\partial S}{\partial x^{a}}\frac{\partial S}{\partial x^{b}}=m^{2}c^{2},
\label{eq:HJ}
\end{equation}
where $S=S(u,v,x^{a})$ is a principal function. We seek a solution of ($\ref{eq:HJ}$) in the form
$S=\tilde{S}(u)+\frac{1}{2}Bv+p_{0a}x^{a}$,
where $B=p_{0}+p_{x}=$\,const and $p_{0a}=$\,const. For a timelike geodesic line we have $B>0$. 
We obtain the principal function given by
\begin{equation}
S=\frac{1}{2B}[(m^{2}c^{2}u-G^{ab}(u,0)p_{0a}p_{0b})]+\frac{1}{2}Bv+p_{0a}x^{a},
\label{eq:S}
\end{equation}
where we define
\begin{equation}
G^{ab}(p,q)=\int_{q}^{q+p}g^{ab}(u)du.
\label{eq:G}
\end{equation}

The components of four-momentum are given by the derivatives of a principal function, 
$p_{i}=\frac{\partial S}{\partial x^{i}}$.
The world line of a particle is determined by Jacobi's theorem,
$\frac{\partial S}{\partial\alpha^{k}}=\beta^{k}$,
where $\alpha^{k}$ are the constants in the principal function ($\ref{eq:S}$), and $\beta^{k}$ are new constants. 
Introducing a finite interval $s$ as a parameter such that
$mc\frac{dx^{a}}{ds}=g^{ab}\frac{\partial S}{\partial x^{b}}$ yields $u=\frac{B}{mc}(s-s_{0})+u_{0}$.
The equations of motion for a massive particle in the field of a plane gravitational wave in terms of $s$ are~\footnote{In agreement with~\cite{eb}.}
\begin{eqnarray}
& & ct(s)=ct(s_{0})+\frac{1}{2}\biggl(\frac{B}{mc}+\frac{mc}{B}\biggr)(s-s_{0}) \nonumber \\ 
& & -\frac{1}{2B^{2}}G^{ab}\biggl(\frac{B}{mc}(s-s_{0}),ct(s_{0})-x(s_{0})\biggr)p_{0a}p_{0b}, \nonumber \\
& & x(s)=x(s_{0})+\frac{1}{2}\biggl(\frac{mc}{B}-\frac{B}{mc}\biggr)(s-s_{0}) \nonumber \\
& & -\frac{1}{2B^{2}}G^{ab}\biggl(\frac{B}{mc}(s-s_{0}),ct(s_{0})-x(s_{0})\biggr)p_{0a}p_{0b}, \nonumber \\
& & x^{a}(s)=x^{a}(s_{0})+\frac{1}{B}G^{ab}\biggl(\frac{B}{mc}(s-s_{0}),ct(s_{0})-x(s_{0})\biggr)p_{0b}.
\label{eq:geo1}
\end{eqnarray}

In a similar manner, we can determine the equations of null geodesic lines in the presence of a plane gravitational wave. The Hamilton-Jacobi equation in this case is equivalent to the eikonal equation,
\begin{equation}
4\frac{\partial \Psi}{\partial u}\frac{\partial \Psi}{\partial v}+g^{ab}\frac{\partial \Psi}{\partial x^{a}}\frac{\partial \Psi}{\partial x^{b}}=0,
\end{equation}
and its solution is of the form
$\Psi=\tilde{\Psi}(u)+\frac{1}{2}Cv+k_{0a}x^{a}$,
where $C=k_{0}+k_{x}=$\,const and $k_{0a}=$\,const. Consequently we find
\begin{equation}
\Psi=-\frac{1}{C}G^{ab}(u,0)k_{0a}k_{0b}+\frac{1}{2}Cv+k_{0a}x^{a},
\end{equation}
with $G^{ab}$ defined in ($\ref{eq:G}$). The components of the four-dimensional wave vector are given by the derivatives of the eikonal, 
$k_{i}=\frac{\partial \Psi}{\partial x^{i}}$.
A world line for a massless particle is again determined from Jacobi's theorem.

Let us define the affine parameter $\pi$ such that
$\frac{dx^{a}}{d\pi}=g^{ab}\frac{\partial \Psi}{\partial x^{b}}$ from which we find $u=C(\pi-\pi_{0})+u_{0}$.
As a result, we obtain the null geodesic lines:
\begin{eqnarray}
& & ct(\pi)=ct(\pi_{0})+\frac{1}{2}C(\pi-\pi_{0}) \nonumber \\
& & -\frac{1}{2C^{2}}G^{ab}(C(\pi-\pi_{0}),ct(\pi_{0})-x(\pi_{0}))k_{0a}k_{0b}, \nonumber \\
& & x(\pi)=x(\pi_{0})-\frac{1}{2}C(\pi-\pi_{0}) \nonumber \\
& & -\frac{1}{2C^{2}}G^{ab}(C(\pi-\pi_{0}),ct(\pi_{0})-x(\pi_{0}))k_{0a}k_{0b}, \nonumber \\
& & x^{a}(\pi)=x^{a}(\pi_{0})+\frac{1}{C}G^{ab}(C(\pi-\pi_{0}),ct(\pi_{0})-x(\tau_{0}))k_{0b},
\label{eq:geo2}
\end{eqnarray}
where $G^{ab}(p,q)$ were defined in ($\ref{eq:G}$).
Formul\ae\,($\ref{eq:geo2}$) are satisfied for all null geodesics, except the curve for which $C=0$. This condition represents a light beam moving parallel to a gravitational wave. In this case, $k_{a}=0$ for an arbitrary $\pi$, and the solution is
\begin{equation}
ct=A(\pi-\pi_{0})+ct_{0},\,\,x=A(\pi-\pi_{0})+x_{0},\,\,x^{a}=x^{a}_{0},
\label{eq:geo3}
\end{equation}
where $A$ is a constant.

\section{A Michelson interferometer in the presence of a plane gravitational wave}

Using the equations of motion for a light beam in the field of a plane gravitational wave, ($\ref{eq:geo2}$) and ($\ref{eq:geo3}$), we will show how to build a simple model of a Michelson interferometer~\cite{11}.
Let us consider three observers at rest:
\begin{eqnarray}
& & \textrm{1.}\,\,\,ct=s-s_{0},\,\,x=0,\,\,y=0,\,\,z=0, \nonumber \\
& & \textrm{2.}\,\,\,ct=s-s_{0},\,\,x=l_{\parallel},\,\,y=0,\,\,z=0, \nonumber \\
& & \textrm{3.}\,\,\,ct=s-s_{0},\,\,x=0,\,\,y=l_{\perp},\,\,z=0.
\end{eqnarray}
Let a light beam leave the world point $(0,0,0,0)$ in the direction of a plane wave, and arrive at the point $({\tilde{s}}_{1}-s_{0},l_{\parallel},0,0)$. In this case equations ($\ref{eq:geo3}$) lead to
\begin{equation}
{\tilde{s}}_{1}-s_{0}=A({\tilde{\pi}}_{1}-{\tilde{\pi}}_{0}),\,\,l_{\parallel}=A({\tilde{\pi}}_{1}-{\tilde{\pi}}_{0}),
\end{equation}
which gives ${\tilde{s}}_{1}-s_{0}=l_{\parallel}$.
Then, let this light beam leave the point $({\tilde{s}}_{1}-s_{0},l_{\parallel},0,0)$ and arrive at the point $({\tilde{s}}_{2}-s_{0},0,0,0)$ (return to the initial space point). Consequently, the equations of motion ($\ref{eq:geo2}$) are
\begin{eqnarray}
& & {\tilde{s}}_{2}-{\tilde{s}}_{1}=\frac{1}{2}C({\tilde{\pi}}_{2}-{\tilde{\pi}}_{1})-\frac{1}{C^{2}}G^{ab}(C({\tilde{s}}_{2}-{\tilde{s}}_{1}),{\tilde{s}}_{1}-l_{\parallel})k_{0a}k_{0b}, \nonumber \\
& & -l_{\parallel}=-\frac{1}{2}C({\tilde{s}}_{2}-{\tilde{s}}_{1})-\frac{1}{2C^{2}}G^{ab}(C({\tilde{s}}_{2}-{\tilde{s}}_{1}),{\tilde{s}}_{1}-l_{\parallel})k_{0a}k_{0b}, \nonumber \\
& & 0=\frac{1}{C}G^{ab}(C({\tilde{s}}_{2}-{\tilde{s}}_{1}),{\tilde{s}}_{1}-l_{\parallel})k_{0b}.
\end{eqnarray}
Therefore ${\tilde{s}}_{2}-s_{0}=2l_{\parallel}$.

Next, let us consider a light beam leaving the point $(0,0,0,0)$ in the direction perpendicular to that of the gravitational wave, and arriving at $(s_{1}-s_{0},0,l_{\perp},0)$. In this case equations ($\ref{eq:geo2}$) give
\begin{eqnarray}
& & s_{1}-s_{0}=\frac{1}{2}C(\pi_{1}-\pi_{0})-\frac{1}{2C^{2}}G^{ab}_{\rightarrow}k_{0a}k_{0b},\,\,l_{\perp}=\frac{1}{C}G^{2b}_{\rightarrow}k_{0b}, \nonumber \\
& & 0=-\frac{1}{2}C(\pi-\pi_{0})-\frac{1}{2C^{2}}G^{ab}_{\rightarrow}k_{0a}k_{0b},\,\, 0=\frac{1}{C}G^{3b}_{\rightarrow}k_{0b},
\label{r}
\end{eqnarray}
where $G^{ab}_{\rightarrow}=G^{ab}(C(\pi_{1}-\pi_{0}),s_{0})=\int_{s_{0}}^{s_{1}}g^{ab}(u)du$.
We obtain the equation for $s_{1}$: 
\begin{equation}
s_{1}-s_{0}=-\frac{l^{2}_{\perp}G^{33}_{\rightarrow}}{\triangle_{\rightarrow}},
\label{eq:dif1}
\end{equation}
where $\triangle_{\rightarrow}=G^{22}_{\rightarrow}G^{33}_{\rightarrow}-(G^{23}_{\rightarrow})^{2}$.

Then, let the same light beam go from the point $(s_{1},0,l_{\perp},0)$ to the initial space point $(s_{2},0,0,0)$. The equations of motion are (with different constants of motion)
\begin{eqnarray}
& & s_{2}-s_{1}=\frac{1}{2}C(\pi_{2}-\pi_{1})-\frac{1}{2C^{2}}G^{ab}_{\leftarrow}k_{0a}k_{0b},\,\,-l_{\perp}=-\frac{1}{C}G^{2b}_{\leftarrow}k_{0b}, \nonumber \\
& & 0=-\frac{1}{2}C(\pi_{2}-\pi_{1})-\frac{1}{2C'^{2}}G^{ab}_{\leftarrow}k_{0a}k_{0b},\,\,0=\frac{1}{C}G^{3b}_{\leftarrow}k_{0b},
\end{eqnarray}
where $G^{ab}_{\leftarrow}=G^{ab}(C(\pi_{2}-\pi_{1}),s_{1})=\int_{s_{1}}^{s_{2}}g^{ab}(u)du$.
In a similar manner we obtain
\begin{equation}
s_{2}-s_{1}=-\frac{l^{2}_{\perp}G^{33}_{\leftarrow}}{\triangle_{\leftarrow}},
\label{eq:dif2}
\end{equation}
where $\triangle_{\leftarrow}=G^{22}_{\leftarrow}G^{33}_{\leftarrow}-(G^{23}_{\leftarrow})^{2}$.

A weak plane gravitational wave in metric (\ref{eq:int}) can be described by the two-dimensional metric tensor
\begin{equation}
g_{ab}=\left[ \begin{array}{cc}
-1+f(u) & h(u) \\
h(u) & -1-f(u) \end{array} \right],
\label{met}
\end{equation}
where $f(u)$ and $h(u)$ are small quantities~\footnote{If we neglect terms quadratic and smaller in $f$ and $h$, then the metric tensor~(\ref{met}) identically satisfies the field equation $R_{uu}=0$.}. The two possible polarizations correspond to setting either quantity to zero. Let us consider a polarization such that $h(u)=0$ and denote it by $h_{+}$. 
From this we find
$\triangle_{\rightarrow}=(s_{1}-s_{0})^{2}+O(A^{2})$,
where $A$ is the amplitude of the gravitational wave. We can linearize equation~($\ref{eq:dif1}$), omitting terms quadratic and smaller in $A$:
\begin{equation}
s_{1}-s_{0}=-\frac{l^{2}_{\perp}}{(s_{1}-s_{0})^{2}}\biggl(-(s_{1}-s_{0})+\int_{s_{0}}^{s_{1}}f(u)du\biggr).
\label{lin}
\end{equation}
The analogous expression can be derived for equation~($\ref{eq:dif2}$).

Let us assume that a plane gravitational wave is monochromatic with wavelength $\lambda$,
$f(u)=A\,cos\frac{2\pi u}{\lambda}$.
We seek the solution of equation~(\ref{lin}) in the form
$s_{1}-s_{0}=l_{\perp}+\triangle s_{\rightarrow}$,
where $\triangle s_{\rightarrow}$ is a small quantity on the order of $A$.
Keeping only linear terms in $\triangle s_{\rightarrow}$ we find
\begin{equation}
\triangle s_{\rightarrow}=-\frac{A\lambda}{2\pi}sin\frac{\pi l_{\perp}}{\lambda}cos\frac{2\pi(s_{0}+l_{\perp}/2)}{\lambda}.
\label{delay}
\end{equation}
Similarly, for the returning beam we have
$s_{2}-s_{1}=l_{\perp}+\triangle s_{\leftarrow}$,
where
\begin{equation}
\triangle s_{\leftarrow}=-\frac{A\lambda}{2\pi}sin\frac{\pi l_{\perp}}{\lambda}cos\frac{2\pi(s_{0}+3l_{\perp}/2)}{\lambda},
\end{equation}
where we used $s_{1}-s_{0}=l_{\perp}$ since this term is on the order of $A$.
Finally, the total delay of the light beam moving perpendicular to the gravitational wave, relative to the parallel beam, is given by
\begin{equation}
\triangle s=\triangle s_{\rightarrow}+\triangle s_{\leftarrow}=-\frac{A\lambda}{2\pi}sin\frac{2\pi l_{\perp}}{\lambda}cos\frac{2\pi(s_{0}+l_{\perp})}{\lambda}.
\label{eq:del}
\end{equation} 

The other polarization (denoted by $h_{\times}$) is described by the metric tensor
\begin{equation}
g_{ab}=\left[ \begin{array}{cc}
-1 & h(u) \\
h(u) & -1 \end{array} \right].
\end{equation}
In this case, equation~($\ref{eq:dif1}$) reduces to
$s_{1}-s_{0}=l_{\perp}$,
whereas for the returning beam we find 
$s_{2}-s_{1}=l_{\perp}$.
For this polarization there is no delay.

Equation~(\ref{eq:del}) is what experimentalists use to detect gravitational waves \cite{bibl,18}. Its relation to incoming gravitational waves with arbitrary propagation directions and polarization states can be found in the literature, e.g.~\cite{lo}. This equation agrees with the corresponding expression in~\cite{fr} if we set the departure moment $s_{0}=0$ and use the relation $\triangle s=\frac{c}{\omega}\triangle\phi$, where $\omega$ is the light beam frequency and $\triangle\phi$ denotes the phase shift. Therefore, treating light as massless particles in the field of a weak gravitational wave and solving the linearized coupled Einstein-Maxwell equations lead to the same result. 

The delay of a light beam depends on the quantity $s_{0}$, which corresponds to the moment when the beam leaves the start point (the time dependence of the gravitational field is fixed by the form of $f(u)$). If the wavelength of a monochromatic gravitational wave is small compared to the size of the interferometer, we can average this delay over $s_{0}$:
$\langle\triangle s\rangle=\frac{1}{\lambda}\int_{0}^{\lambda}\triangle s(s_{0})ds_{0}$.
The physical reason for averaging arises from the fact that we do not know when the front of a gravitational wave hits the point of the departure of the beam. However, as for each periodic function, such an expression vanishes and is of no interest.

\section{Light beam delay in the second approximation}

In this section we will generalize the results of the preceding section and consider quadratic terms. Let the two-dimensional metric tensor $g_{ab}$ of a weak gravitational wave be given by
\begin{equation}
g_{ab}= \left[ \begin{array}{cc}
-1+f^{(1)} & h^{(1)} \\
h^{(1)} & -1+g^{(1)} \end{array} \right],
\end{equation}
where $f^{(1)},g^{(1)},h^{(1)}$ are small quantities on the same order. 
In the linear approximation $f^{(1)}+g^{(1)}=O(A^{2})$, where $A$ is the amplitude of such a wave on the order of $f^{(1)}$. Therefore, without loss of generality we may assume
$f^{(1)}+g^{(1)}=g^{(2)}$.
The field equation is obtained from $R_{uu}=0$:
\begin{equation}
\frac{1}{2}({\dot{f}}^{(1)2}+2f^{(1)}{\ddot{f}}^{(1)}+{\ddot{g}}^{(2)}+{\dot{h}}^{(1)2}+2h^{(1)}{\ddot{h}}^{(1)})=0,
\label{eq:field}
\end{equation}
which gives $g^{(2)}$ as a function of $f^{(1)}$ and $h^{(1)}$.

If we consider a polarization $h_{+}$ with $h^{(1)}=0$ and put $s_{1}-s_{0}=l_{\perp}$ in terms on the order of $A^{2}$, then equation~($\ref{eq:dif1}$) becomes
\begin{equation}
(s_{1}-s_{0})^{3}-l_{\perp}^{2}(s_{1}-s_{0})+l_{\perp}^{2}\int_{s_{0}}^{s_{1}}(f^{(1)}+f^{(1)2})du-l_{\perp}\biggl(\int_{s_{0}}^{s_{1}}f^{(1)}du\biggr)^{2}=0,
\label{bd}
\end{equation}
Note that this equation does not contain the quantity $g^{(2)}$.
Let us apply the above results to the case of a monochromatic plane wave,
$f^{(1)}=A\,cos\frac{2\pi u}{\lambda}$.
For such a wave, equation~(\ref{eq:field}) leads to
$g^{(2)}=\frac{4\pi^{2}A^{2}}{\lambda^{2}}(3cos^{2}\frac{2\pi u}{\lambda}-1)$,
and this quantity clearly is not a periodic function of $u$~\footnote{We could assume $f^{(1)}+g^{(1)}=f^{(2)}$, but then equation.~(\ref{bd}) would contain the nonperiodic function $f^{(2)}$. Thus, it would be impossible to average the light beam delay over $s_{0}$ (the moment when the beam leaves the start point) independently of the initial conditions.}.
We seek the solution in the form
$s_{1}-s_{0}=l_{\perp}+\triangle s_{\rightarrow}$,
where $\triangle s_{\rightarrow}$ is again a small quantity on the same order as $A$. Thus, up to quadratic terms in $\triangle s_{\rightarrow}$ we obtain
\begin{eqnarray}
& & \triangle s_{\rightarrow}=-\frac{A\lambda}{2\pi}sin\frac{\pi l_{\perp}}{\lambda}cos\frac{\pi}{\lambda}(2s_{0}+l_{\perp})-\frac{A^{2}\lambda}{16\pi}sin\frac{2\pi l_{\perp}}{\lambda}cos\frac{2\pi}{\lambda}(2s_{0}+l_{\perp}) \nonumber \\
& & -\frac{A^{2}l_{\perp}}{4}+\frac{A^{2}\lambda^{2}}{8\pi^{2}l_{\perp}}\biggl(sin\frac{\pi l_{\perp}}{\lambda}cos\frac{\pi}{\lambda}(2s_{0}+l_{\perp})\biggr)^{2} \nonumber \\
& & +\frac{A^{2}\lambda}{4\pi}sin\frac{\pi l_{\perp}}{\lambda}cos\frac{\pi}{\lambda}(2s_{0}+l_{\perp})cos\frac{2\pi}{\lambda}(s_{0}+l_{\perp}).
\label{eq:31}
\end{eqnarray}
The above expression contains the quadratic corrections to equation~(\ref{delay}). The mean value of this quantity over the moment of the departure of the beam $s_{0}$,
$\langle\triangle s_{\rightarrow}\rangle=\frac{1}{\lambda}\int_{0}^{\lambda}\triangle s_{\rightarrow}(s_{0})ds_{0}$,
is now different from zero:
\begin{equation}
\langle\triangle s_{\rightarrow}\rangle=\frac{A^{2}\lambda^{2}}{16\pi^{2}l_{\perp}}sin^{2}\frac{\pi l_{\perp}}{\lambda}-\frac{A^{2}l_{\perp}}{4}+\frac{A^{2}\lambda}{16\pi}sin\frac{2\pi l_{\perp}}{\lambda}.
\end{equation}
We remind that the moment $s_{0}$ is relative to the instant when a front of the gravitational wave in question hits the departure point, and we average with respect to $s_{0}$ since we do not know the value of this quantity.

\begin{figure}[t]
\centering
\includegraphics[width=2.5in]{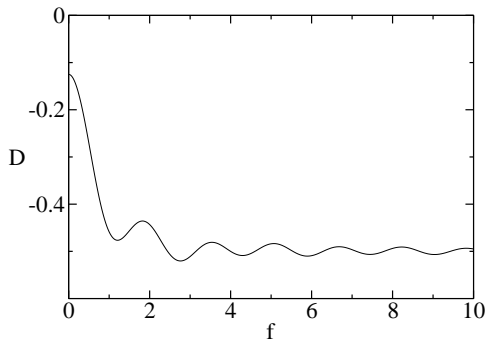}
\caption{\label{pic1} The normalized average light beam delay $D$ as a function of the normalized wave frequency $f$ for the polarization $h_{+}$. }
\end{figure}

For the returning beam we must replace $s_{1}$ with $s_{2}$ and $s_{0}$ with $s_{1}$ (since the beam departs from the reflection point right after it arrives there), which gives
$s_{2}-s_{1}=l_{\perp}+\triangle s_{\leftarrow}$,
where $\triangle s_{\leftarrow}$ is described by equation~($\ref{eq:31}$) in which $s_{0}$ is replaced with $s_{0}+l_{\perp}+\triangle s_{\rightarrow}$. The result is
\begin{equation} 
\langle\triangle s_{\leftarrow}\rangle=\langle\triangle s_{\rightarrow}\rangle-\frac{A^{2}\lambda}{4\pi}sin^{2}\frac{\pi l_{\perp}}{\lambda}sin\frac{2\pi l_{\perp}}{\lambda}.
\label{di}
\end{equation}
Therefore, the total average delay of the light beam is given by
\begin{eqnarray}
& & \langle\triangle s\rangle=\frac{A^{2}\lambda^{2}}{8\pi^{2}l_{\perp}}sin^{2}\frac{\pi l_{\perp}}{\lambda}-\frac{A^{2}l_{\perp}}{2}+\frac{A^{2}\lambda}{8\pi}sin\frac{2\pi l_{\perp}}{\lambda} \nonumber \\
& & -\frac{A^{2}\lambda}{4\pi}sin^{2}\frac{\pi l_{\perp}}{\lambda}sin\frac{2\pi l_{\perp}}{\lambda}.
\end{eqnarray}
We observe that $\langle\triangle s\rangle$ is negative for all values of $\lambda$. If $l_{\parallel}=l_{\perp}$, then the light beam moving perpendicular to the direction of a gravitational wave lags (on average with respect to the departure moment $s_{0}$) behind the one which moves parallel, provided both beams left the light source at the same time. The dependence of the quantity
$D=\frac{\langle\triangle s\rangle}{A^{2}l_{\perp}}$
on the variable
$f=\frac{\pi l_{\perp}}{\lambda}$
is shown in figure~\ref{pic1}.

\begin{figure}[t]
\centering
\includegraphics[width=2.5in]{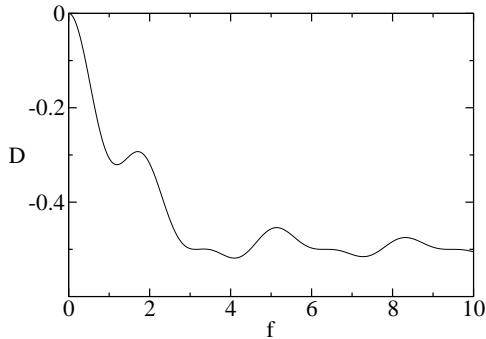}
\caption{\label{pic2} The normalized average light beam delay $D$ as a function of the normalized wave frequency $f$ for the polarization $h_{\times}$. }
\end{figure}

Now we consider the other polarization $h_{\times}$ for which $f^{(1)}=0$. 
Equation~($\ref{bd}$) is replaced by 
\begin{equation}
(s_{1}-s_{0})^{3}-l_{\perp}^{2}(s_{1}-s_{0})+l_{\perp}^{2}\int_{s_{0}}^{s_{1}}(h^{(1)2})du-l_{\perp}\biggl(\int_{s_{0}}^{s_{1}}h^{(1)}du\biggr)^{2}=0.
\end{equation}
For a monochromatic plane wave,
$h^{(1)}=A\,cos\frac{2\pi u}{\lambda}$,
instead of equation~(\ref{eq:31}) we find
\begin{eqnarray}
& & \triangle s_{\rightarrow}=\frac{A^{2}\lambda^{2}}{2\pi^{2}l_{\perp}}\biggl(sin\frac{\pi l_{\perp}}{\lambda}cos\frac{\pi}{\lambda}(2s_{0}+l_{\perp})\biggr)^{2}-\frac{A^{2}l_{\perp}}{4} \nonumber \\
& & -\frac{A^{2}\lambda}{16\pi}sin\frac{2\pi l_{\perp}}{\lambda}cos\frac{2\pi}{\lambda}(2s_{0}+l_{\perp}).
\end{eqnarray}
The returning beam again satisfies formula (\ref{di}). Averaging over the departure moment $s_{0}$ gives
\begin{equation}
\langle\triangle s\rangle=\frac{A^{2}\lambda^{2}}{2\pi^{2}l_{\perp}}sin^{2}\frac{\pi l_{\perp}}{\lambda}-\frac{A^{2}l_{\perp}}{2}-\frac{A^{2}\lambda}{4\pi}sin^{2}\frac{\pi l_{\perp}}{\lambda}sin\frac{2\pi l_{\perp}}{\lambda}.
\end{equation}
We see that $\langle\triangle s\rangle$ is negative for all values of the length of a gravitational wave, and tends asymptotically to zero as $\lambda\rightarrow\infty$. In figure~\ref{pic2} we show the function $D(f)$. Note that for both polarizations $D\rightarrow1/2$ as $\lambda\rightarrow0$.

From figures \ref{pic1} and \ref{pic2} we conclude that a one-to-one relation between the average light beam delay and the wavelength exists only for waves longer than $\sim\pi l_{\perp}$. Moreover, the precision of the measurement of the delay is proportional to the length of the arms of the interferometer. Therefore, if we try to increase the precision by increasing this length, we end up with decrease of the range of waves measurable in this way.

\section{Summary}

In this work we showed that the time delay of a light beam in a Michelson interferometer, obtained by treating light as massless particles moving on geodesics, agrees with the corresponding solution of the linearized coupled Einstein-Maxwell equations. We also calculated this delay in the second (quadratic) approximation. We considered the simple case of a plane wave with the direction parallel to one arm of the interferometer, and treated the two polarizations separately. However in reality, one should expect a broad spectrum of frequencies and a combination of both polarizations. 

Averaging the light beam delay $\triangle s$ over the moment of the departure of the beam $s_{0}$ would make sense if either wavelength $\lambda$ is much smaller than the arm length $l_{\perp}$ or we measure $\langle\triangle s\rangle$ many times. Since we excluded the former, this method would not work in the case of supernova explosions that radiate significantly for a few seconds only. Instead, the presented model could be used (in principle) for continuous sources of gravitational waves such as stellar binaries containing a neutron star. As we stated in section~1, second-order effects in gravitational-wave detection are extremely small and thus irrelevant experimentally. However, they might be of some academic interest.

\begin{center}
{\bf Acknowledgment}
\end{center}
The author would like to express his gratitude to Professor S.~L.~Ba\.{z}a\'{n}ski for supervision of the M.S. thesis on which this paper is based (Faculty of Physics, University of Warsaw, Poland, 1999).

\end{document}